\documentclass[preprint, 1p, 12pt]{elsarticle}
\pdfoutput=1

\usepackage{graphicx}
\usepackage{dcolumn}
\usepackage{bm}
\usepackage{epsfig}
\usepackage{float} 
\usepackage{amsmath}
\usepackage{amssymb}
\usepackage{subfig}
\usepackage{pifont}
\usepackage{natbib}

\biboptions{sort&compress}

\begin{document}

\begin{frontmatter}

\title{Verifying Spent Fuel Containers Before Deep Geological Storage with Cosmic Ray Muons}

\author[LANL,UNM]{D. Poulson}
\author[LANL]{J. M. Durham}
\author[LANL]{J. D. Bacon}
\author[LANL]{E. Guardincerri}
\author[LANL]{C. L. Morris}

\address[LANL]{Los Alamos National Laboratory, Los Alamos, NM 87545, USA}
\address[UNM]{University of New Mexico, Albuquerque, NM 87131, USA}

\begin{abstract}
International nuclear safeguards inspectors do not have a method to verify the contents of sealed storage casks containing spent reactor fuel.  The heavy shielding that is used to limit radiation emission attenuates and scatters photons and neutrons emitted by the fuel, and thereby hinders inspection with these probes.  This problem is especially pressing given the policy decisions of several nations to begin permanent disposal of spent fuel in deep geological repositories.  Radiography with cosmic-ray muons provides a potential solution, as muons are able to penetrate the cask and fuel and provide information on the cask contents.  Here we show in simulation that muon scattering radiography can be used to inspect the contents of sealed geological storage casks, and can discern between a variety of plausible diversion scenarios.  This technique can be applied immediately prior to permanent interment in a geological repository, giving inspectors a final opportunity to verify State declarations of spent fuel disposal.
\end{abstract}


\end{frontmatter}

\section{Introduction}

The lack of options for the final disposal of highly radioactive spent nuclear fuel remains a challenge to the nuclear power industry.  Multiple countries are now considering permanent disposal of waste deep underground in geological repositories.  Finland and Sweden will open the first geological repositories for high-level waste, which are expected to begin accepting spent reactor fuel assemblies in the 2020s and 2030s, respectively \cite{Finland,Sweden,EuropeRepository}.  The spent fuel will be encapsulated within heavily shielded copper and iron containers \cite{p24}, and then transferred to these facilities for permanent underground interment.

In most nations, fissile material is monitored by the International Atomic Energy Agency (IAEA), which has the responsibility under the Treaty on the Nonproliferation of Nuclear Weapons to implement nuclear safeguards in non-nuclear-weapons states \cite{NPT}.  Spent nuclear fuel rods contain plutonium that is created via neutron capture on uranium nuclei during reactor operations, which is categorized as irradiated direct use material by the IAEA \cite{IAEA_SG_glossary} and thereby falls under safeguards.  The IAEA monitors and inventories fuel through all steps of the nuclear fuel cycle, including final disposal.  Safeguards are terminated when nuclear material is ``practically irrecoverable," subject to agreements between the IAEA and the State under safeguards \cite{INFIRC153}.  Spent fuel that is buried in a repository may meet that criterion.  Therefore, a final verification after encapsulation and immediately preceding burial is the last chance for inspectors to evaluate state declarations of spent fuel disposal.

Previous studies on verifying the content of metallic dry storage casks via measurements of neutrons and photons which escape the cask shielding has shown that these probes cannot provide sufficient information to determine if fuel has been removed from the center of the cask \cite{Ziock, INL_Doel, AggiesNeutronsDontWork, Swinhoe}.  More recent work has shown that the neutron deficit, following the removal of low-burnup bundles, can be obscured by replacing some of them with high-burnup fuel with a larger neutron emission rate \cite{INMM_NeutronsDontWork}. 
Recognizing this unsolved problem in nuclear safeguards, the IAEA 2018 Long Term R\&D plan lists the development of ``safeguards equipment to establish and maintain knowledge of spent fuel in shielding/storage/transport containers" as a Top Priority R\&D need \cite{STR385}.

Cosmic-ray muon scattering radiography presents an alternate method for investigating cask contents.  Muons are a highly penetrating and passive probe that is capable of penetrating storage cask's shielding and interrogating the fuel assemblies inside.  There have been explorations of various nuclear waste assessment scenarios via simulation \cite{Gustafsson, Jonkmans, Furlan, Liao, Clarkson1, Ambrosino, Clarkson2, CHATZI1, Frazao, Chatzi2, Poulson2017, Boniface, Checchia, Liu}, and a recent statistics-limited measurement proved that cosmic-ray muon scattering measurements are sensitive to removal of fuel assemblies from a sealed dry storage cask \cite{Durham2}.    

In this paper, we discuss simulations of cosmic ray muon scattering in a storage cask containing spent reactor fuel, using a realistic cask design that will be used in the Finnish and Swedish repositories, realistic fuel designs, and a simulated detector with realistic angular resolution. We consider several different diversion scenarios:  a single missing fuel assembly, half a missing fuel assembly, and replacement of a spent fuel assembly with a dummy assembly made of lead.  This work shows that a muon tracking detector with reasonable angular resolution can distinguish these diversion scenarios from a completely full storage cask with high confidence by recording 24 hours of cosmic-ray muon data.  Such a system could be built using existing technology, and could potentially provide a method for final verification of encapsulated spent fuel before permanent disposal.

\section{Simulation Setup}
\begin{figure}[htbp]
\centering
\includegraphics[width = \textwidth]{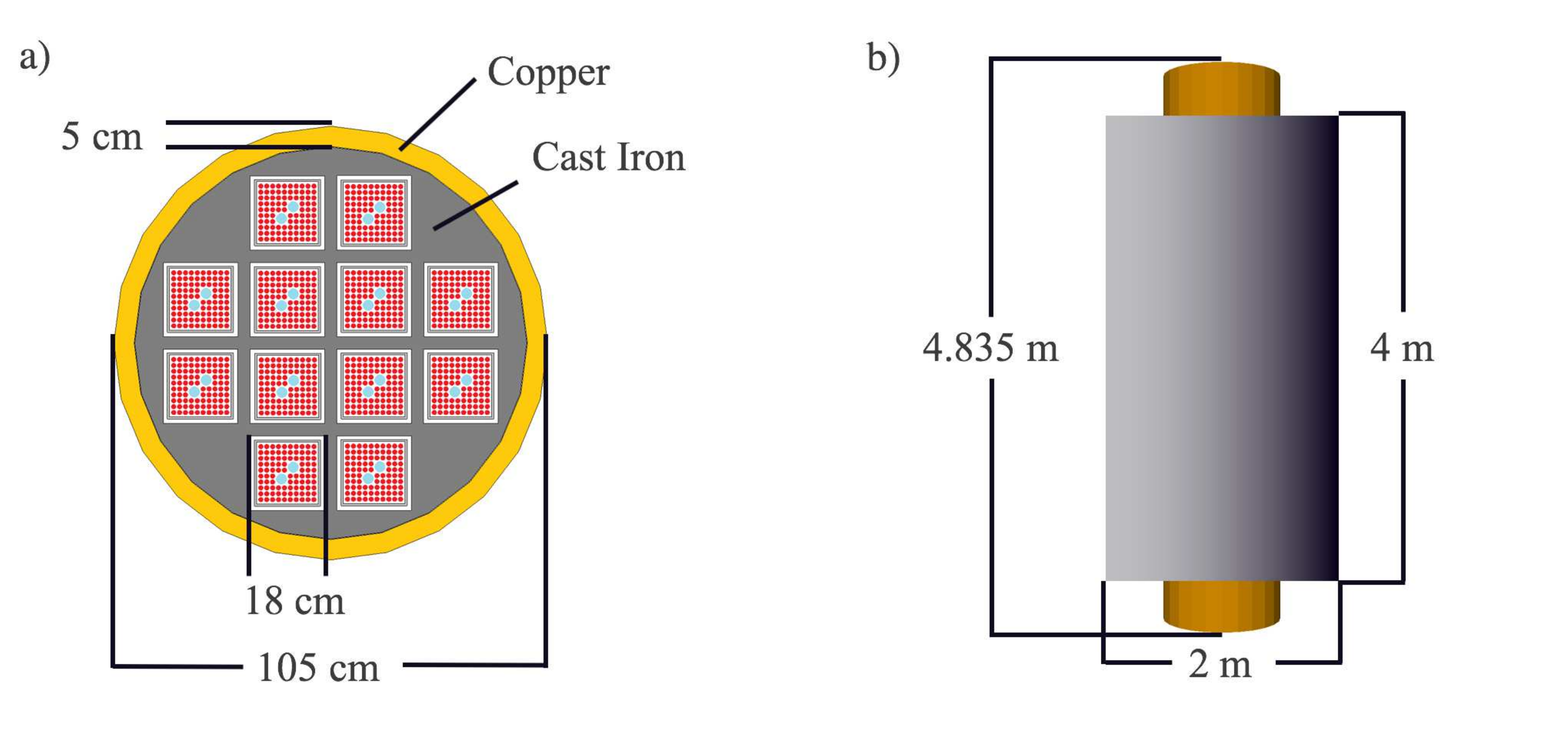}
\caption{a) Cross-section of the simulated storage cask filled with $10\times 10$ BWR fuel bundles. The red circles represent fuel rods and the blue circles represent water rods. b) Top-down view of the simulated detector plane over the cask.}
\label{fig:CC_CrossSection}
\end{figure}

The GEANT4 toolkit \cite{p25} is used to simulate muon passage through fuel casks. The simulated cask geometry follows the description of the actual casks that will be used in the Finnish and Swedish repositories found in \cite{p24}, which includes specifications for both pressurized water reactor (PWR) and boiling water reactor (BWR) fuel casks.  Both cask designs are cylinders composed of an interior cast iron insert with slots to accommodate the fuel assemblies, surrounded by a copper shell to act as a barrier against corrosion. The cast iron insert has a diameter of 95 cm and the copper shell has a thickness of 5 cm, giving the cask an outer diameter of 1.05 m.  The cask length is 4.835 m. The insert for PWR assemblies has four $26 \times 26$ cm$^2$ slots spaced in an array with 37 cm separation from center to center. Inserts for BWR assemblies have twelve $18\times18$ cm$^2$ slots spaced in an array with 21 cm separation from center to center. A cross section of the BWR cask with fuel is shown in Fig. \ref{fig:CC_CrossSection}a.  In this paper, we consider the BWR cask, as the reduced fuel size and larger number of fuel bundles in the cask (relative to the PWR cask) make this a more challenging verification scenario.  It is reasonable to assume that the analysis discussed here would also apply to PWR casks, where the number of fuel bundle assemblies is reduced and their size is increased.

The simulated BWR fuel assemblies are $10\times10$ bundles of individual fuel rods with a cross-sectional area of $16\times 16$ cm$^2$. Each assembly contains 92 UO$_2$ fuel rods and two water rods. The fuel rods have a radius of 4.775 mm and an outer cladding layer of zirconium alloy that is 0.71 mm thick. The water rods are composed of a hollow zirconium alloy tube, 0.71 mm thick with an outer radius of 12.725 mm. The $10\times10$ array of fuel rods are encased in a square 10 mm thick aluminum shell. The assemblies do not vary along their 3.85 m length. All fuel assemblies were placed inside of the fuel cask with identical orientation, as a 90$^\circ$ rotation of a fuel bundle has no effect on the muon scattering signal.

\begin{figure}[htbp]
\centering
\includegraphics[width = \textwidth]{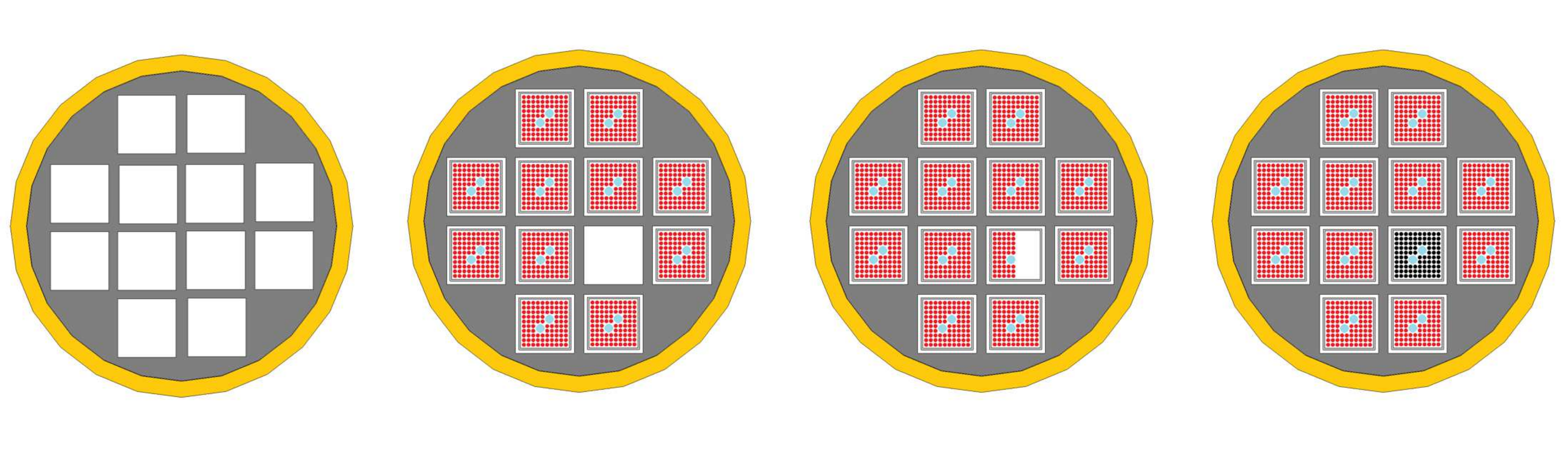}
\caption{The diversion scenarios considered in this study.  From left to right: a completely empty cask, a cask with one missing assembly, a cask with half of one assembly missing, and a cask where an assembly has been replaced by a dummy assembly containing lead in place of UO$_2$ fuel.}
\label{fig:Diversions}
\end{figure}

The BWR fuel cask simulations are broken in to five scenarios: a full cask (nominal scenario), an empty cask, a cask with one missing bundle, a cask with one half missing bundle (half of the fuel rods removed from one side), and a cask with one fuel bundle replaced with lead (UO$_2$ replaced with Pb rods), see Fig. \ref{fig:Diversions}. For each scenario, two horizontal detector planes are used to simulate muon tracking detectors above and below the cask, with the cask placed horizontally between them. The detector planes are aligned horizontally and separated vertically by 1.5 m. Each detector has dimensions of 4 meters along the length of the cask and 2 meters along its width, see Fig. \ref{fig:CC_CrossSection}b. 

The momentum and angular distributions of the simulated muons are randomly sampled from Reyna's parameterized distribution described in \cite{p26}.  Only muons which pass through both detectors and the cask or within 5 cm of its edge were put through the full simulation chain. In order to include realistic detector resolution effects, each muon's incoming and outgoing directional vectors were given a Gaussian blur, giving the detector a limiting resolution of 10 milliradians on the muon scattering angle. Each data set consisted of 2.33 $\times 10^7$ generated muons, which corresponds to approximately 12 hours counting time for this detector geometry. 

\section{Analysis and Discussion}

Here we describe two different analysis techniques applied to the simulated data.  First, we perform a straightforward analysis comparing the muon scattering angles in a full cask with the various diversion scenarios, using 24 hours worth of simulated data on a static cask.  In the second technique, we consider two 12 hour data sets, where the cask is rotated 90$^\circ$ about its central axis in between measurements.  From this, we construct a sinogram with limited angular coverage and use it to constrain a model of the cask contents.

\subsection{Scattering Angle Analysis}

A virtual tally plane is created at the center of the cask.  The plane has a width of 1.2 m and length of 3 m, and is divided into 2 cm wide pixels that run the length of the plane.  Since each fuel cask is uniform along its long axis along within the length of the fuel bundles, segmentation along this direction is not required.  Each pixel has a corresponding histogram for recording muon scattering angles. 

The array of BWR fuel bundles inside the cask is divided into columns across the width of the cask, with the two outer columns having two fuel assemblies, and the two inner bundles having four fuel assemblies in the nominal case of a full cask.  Muon tracks in the upper simulated detector are given a Gaussian blur and then projected to the center of the cask. If the muon intersects the virtual tally plane, and does not cross between fuel bundle columns, the corresponding histogram of the intersected pixel is filled with the scattering angle, which is measured by comparing the muon's blurred incident and blurred exiting tracks. 

\begin{figure}[htbp]
\centering
\includegraphics[width = \textwidth]{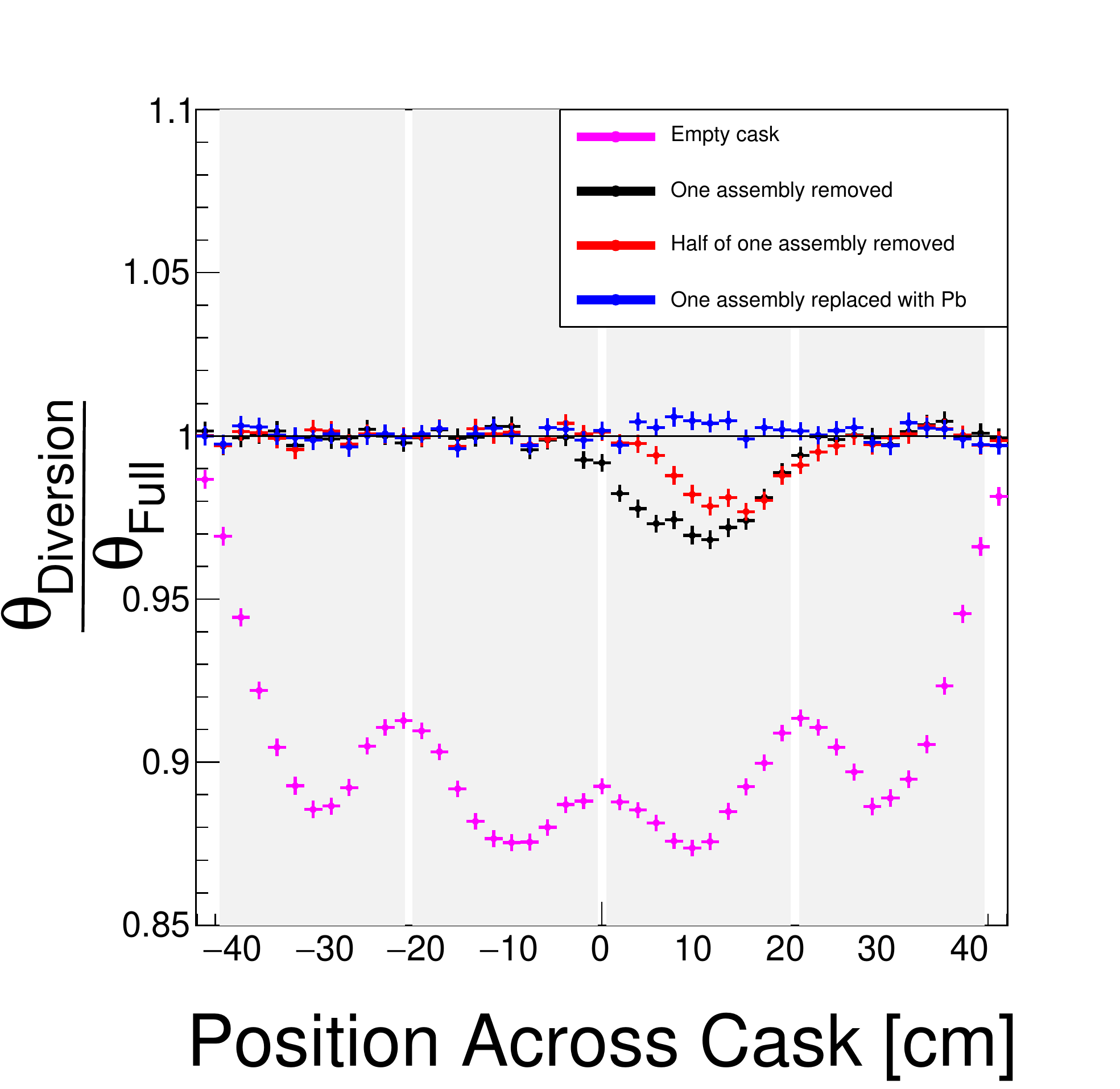}
\caption{The ratio of $\theta_\text{scat}$ for each diversion scenario to that of a full cask, for the various diversion scenarios. The shaded regions indicate the locations of fuel bundle columns (containing 2, 4, 4, and 2 fuel assemblies form left to right).}
\label{fig:CC_RatioFullDay}
\end{figure}

The average value of the muon scattering angle in each histogram ($\theta_\text{scat}$) is calculated for each pixel, for each simulated scenario. The ratio of $\theta_\text{scat}$ of each diversion scenario to that of a full fuel cask along the width of the cask is plotted in Fig. \ref{fig:CC_RatioFullDay} for 24 hrs counting time (using two independent 12 hr data sets). The horizontal lines indicate the 2 cm width of each pixel and the vertical lines show the statistical uncertainty of the calculated ratio. A black line at unity is shown to represent the expectation of comparisons with a full cask. 

The muon scattering ratio for a completely empty cask is obviously different from the expected ratio for a full cask by up to 48 $\sigma$. For the diversion scenarios, the muon scattering ratio is consistent with the expected ratio for the first column. The second and fourth columns also show muon scattering ratios that are consistent with the expected ratio, save for the few centimeters closest to the third column. In the third column, where the defect in the cask loading is located, the divergence of the muon scattering ratio for a single fuel bundle removed reaches 11.2$\sigma$ away from that expected for a full cask, while the ratio for a half bundle removed reaches 8.3$\sigma$. The muon scattering ratio for replacement of a fuel bundle with lead is slightly above one, due to the shorter radiation length $X_0$ for lead than UO$_2$, which are 0.56 cm and 0.61 cm, respectively.  However, the significance of this signal is under 2$\sigma$ in each individual bin from the expected ratio for a full cask, precluding definitive statements about replacement in this measurement scenario.

\subsection{Model Fitting Analysis}

\begin{figure}[htbp]
\centering
\includegraphics[width = \textwidth]{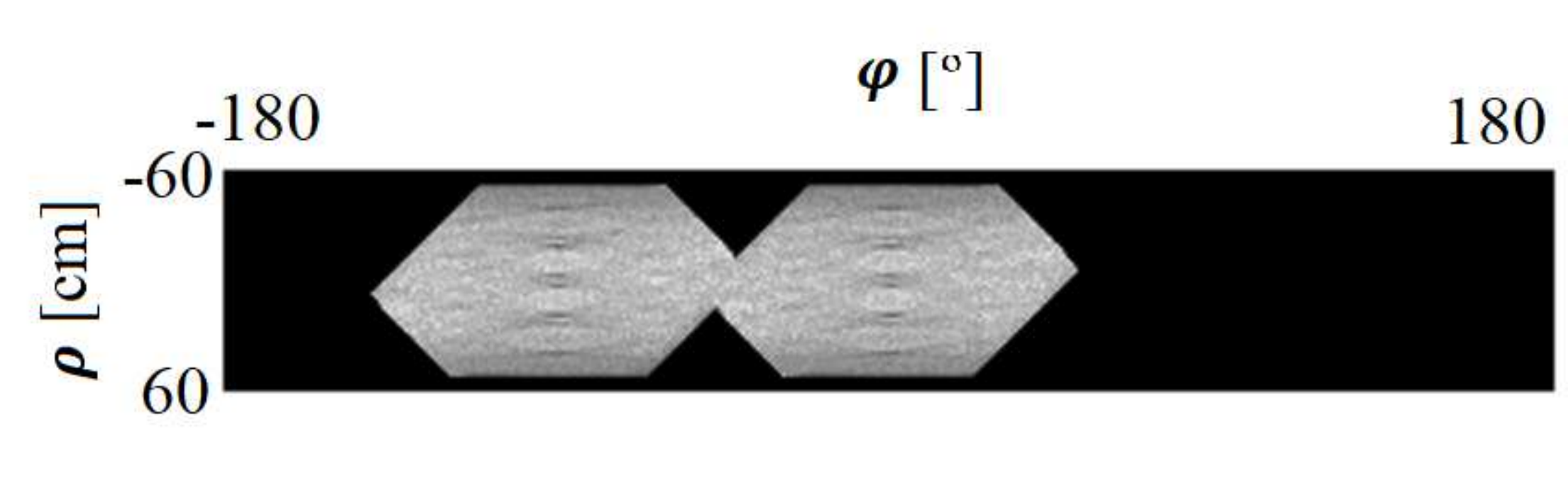}
\caption{A partial sinogram of a full fuel cask created from two 12 hour counting time data sets, separated by a 90$^\circ$ rotation. The sinograms had 60, 2 cm divisions along the cask width ($\rho$) and 360, 1$^\circ$ divisions in rotation about the central axis of the cask ($\varphi$). Black indicates regions where no scattering data was collected.}
\label{fig:CC_Sinogram}
\end{figure}

A second analysis is performed to investigate the sensitivity in the muon scattering signal to the replacement of spent fuel with lead, using the model fitting techniques first described in \cite{Poulson2019}, but with limited angular coverage. Additional simulated data sets representing 12 hours of counting time are produced with the cask rotated 90$^\circ$ about its central axis. The simulated data set for one 12 hr counting period in the initial configuration was combined with one 12 hr counting period after rotation, creating a 24 hr data set. Rather than projecting the muon tracks to a single tally plane at the cask's center, each track was projected to hit a tally plane to which the incident muon track was approximately normal. Each tally plane also passed through the cask's center, with 1$^\circ$ offsets between them and pixel divisions of 2 cm along the width of the cask. The corresponding scattering histograms for each pixel were fit with the multigroup method 
(see \cite{Morris2012} for details), which provides an estimate of the number of radiation lengths that the muons in the histogram had traversed. This information is used to construct a sinogram of the fuel cask, shown in Fig. \ref{fig:CC_Sinogram}, as described in \cite{Poulson2017}.  Each sinogram element represents the number of radiation lengths that muons, with an offset $\rho$ from the cask's center and a rotation $\varphi$ about the cask's central axis, pass through when traversing in the cask.  Here, due to the fact that muons are sampled in a limited range of zenith angles set by the detector geometry, there is only partial angular coverage of the fuel cask.

\begin{figure}[htbp]
\centering
\includegraphics[width = \textwidth]{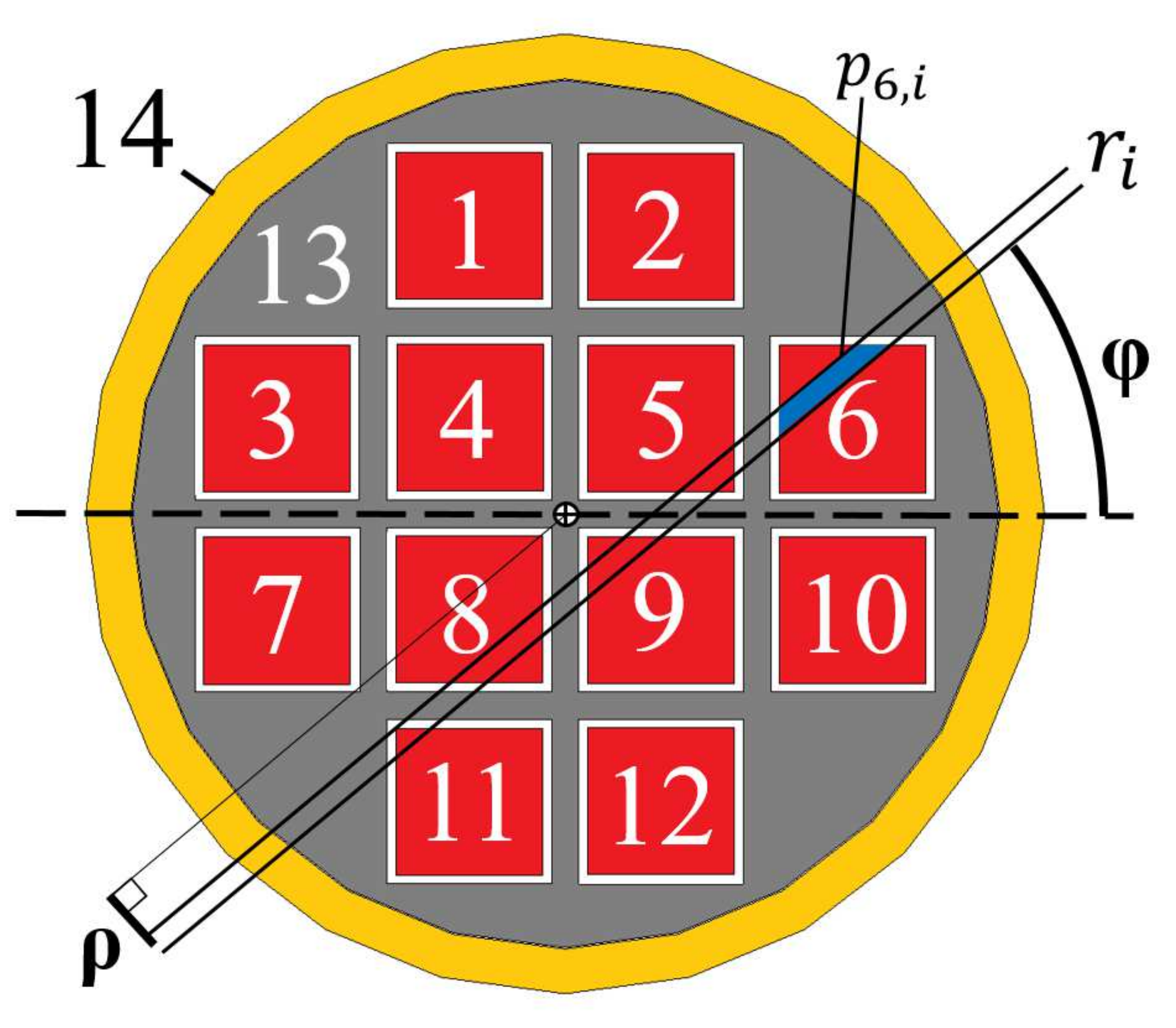}
\caption{The BWR cask model with the numbered model elements and ray $r_i$ passing through at angle $\varphi$ with offset $\rho$ from the cask center.}
\label{fig:CC_Model}
\end{figure}

We define a model of the cask using its component parts: the copper shell, the cast iron insert, and the 12 fuel assemblies that should be present in a full cask.   In this simplified cask model, each element in the partially filled sinogram can be represented as 
\begin{align}
    s_i = \sum_{k \in r_i} p_{ki} a_k,
\end{align}
where $s_i$ is the value in the sinogram defined by ray $r_i$, $p_{ki}$ is the path length of $r_i$ through model element $k$, and $a_k$ is the scattering density of the cask model element $k$. Fig. \ref{fig:CC_Model} shows the geometry of the cask model with a ray corresponding to a sinogram element.

 The goal is now to find the individual values $a_k$ corresponding to each of the 14 cask elements, which are numbered as shown in Fig. \ref{fig:CC_Model}. From these definitions, a $\chi^2$ can be defined as
 \begin{align}
     \chi^2 \equiv \sum_{i} \left( \sum_{j} p_{ji} a_j - s_i \right)^2.
 \end{align}
The optimal values of $a_k$ are found where this $\chi^2$ is minimized, i.e. by solving 
\begin{align}
    \frac{\partial }{\partial a_k} \chi^2 = 0 = 2\sum_i\left( \sum_{j} p_{ji} a_j - s_i \right)p_{ki}.
\end{align}
This expression can be rearranged and represented by the matrix equation
\begin{align}
    \bm{s} = \bm{P}\bm{a},
\end{align}
where $\bm{s}_k = \sum\limits_{i} p_{ki} s_i$, $\bm{P}_{kj} = \sum\limits_{i} p_{ki}p_{ji}$, and $\bm{a}_k = a_k$. The solution 
\begin{align}
    \bm{a} = \bm{P}^{-1}\bm{s},
    \label{eq:CC_Inv}
\end{align}
is found via matrix inversion. 

\begin{figure}[htbp]
\centering
\includegraphics[width = \textwidth]{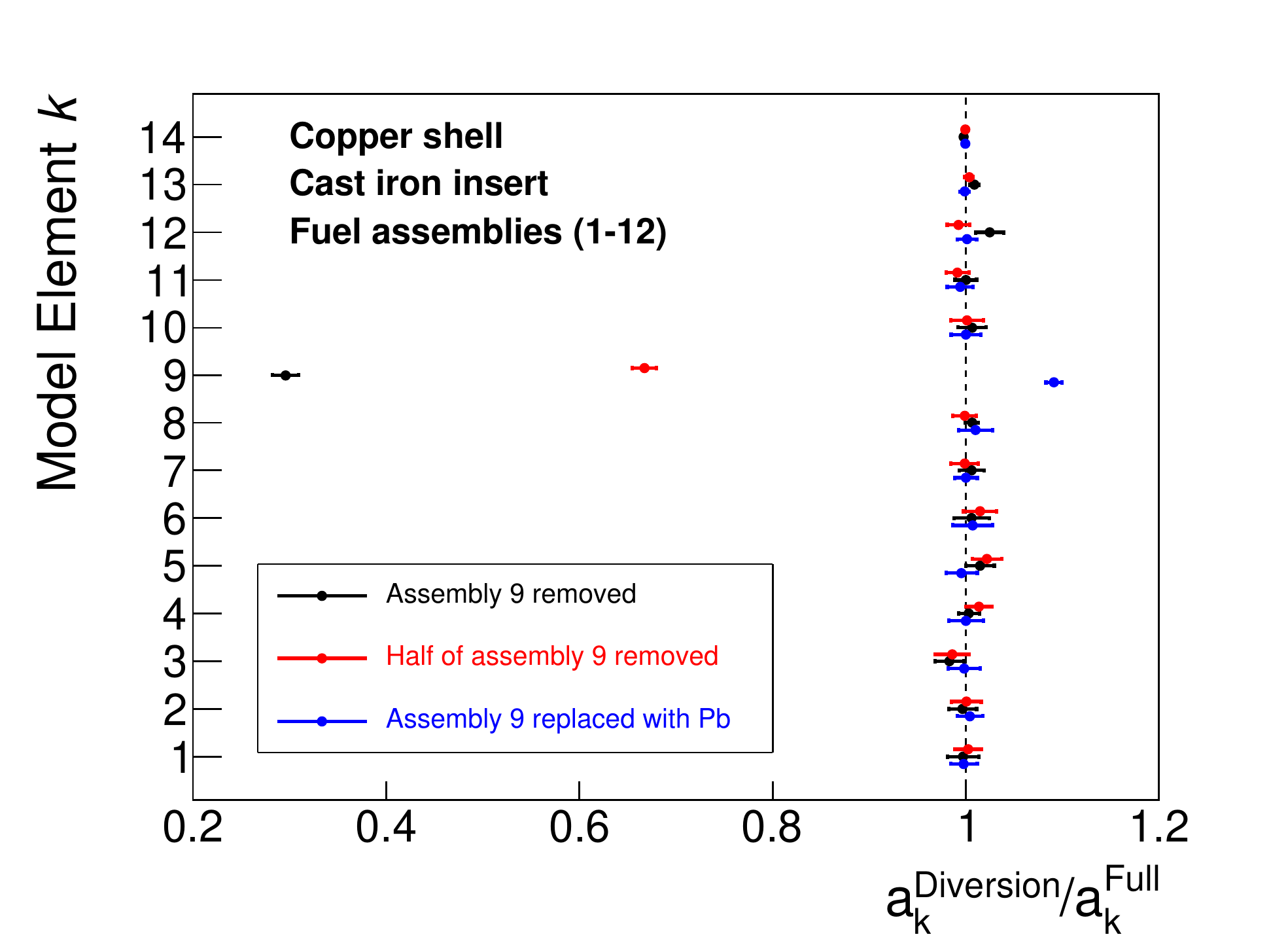}
\caption{The calculated ratio of scattering densities for each model element, for the various diversion scenarios.  The data points are offset slightly along the vertical axis for clarity.}
\label{fig:aks}
\end{figure}

The scattering densities $a_{k}$ of each cask model element $k$ are calculated from the partial sinograms for the full cask and for each diversion scenario. We compare the scattering densities in each diversion scenario $a_{k}^{Diversion}$ to those found for a full cask $a_{k}^{Full}$ by showing their ratio in Fig. \ref{fig:aks}. A statistical uncertainty is estimated by repeating the simulation and analysis with 10 independent simulation samples in each scenario; the standard deviation on the ratio $a_{k}^{Diversion}/a_{k}^{Full}$ found by comparing these data sets is shown as the error bar on the points in Fig. \ref{fig:aks}.  

The cast iron insert and copper shell (model elements 13 and 14, respectively) are well constrained by the data, as evidenced by their relatively small uncertainties since all muons that pass through the cask must pass through these elements. One missing fuel bundle deviates from the expected ratio by 102$\sigma$. For one half fuel bundle removed, the ratio deviates from the expected value of one by 28$\sigma$.  One missing fuel bundle replaced with Pb rods deviates from the expected ratio by 5.1 $\sigma$.  This demonstrates that these diversion scenarios can be detected with 24 hours of muon scattering data.

\section{Summary}

There is a pressing need in international nuclear safeguards for a method that is able to verify the contents of spent fuel casks before they are interred in geological storage.  Cosmic-ray muons provide a highly penetrating probe than can access encapsulated fuel through shielding and emerge with useful radiographic information.  Here we have shown that 24 hours of measurements of cosmic ray muon scattering in a geological storage cask are sufficient to show if a fuel assembly has been removed, or replaced by a dummy assembly made out of lead.  We have used realistic models of the storage cask, the nuclear fuel, and have applied angular resolution smearing to the muon tracks to represent resolution in a detector that can be built with existing technology.  Such an instrument could conceivably be built and used to solve this problem in international nuclear safeguards.

\section{Acknowledgements}

We thank the Los Alamos National Laboratory's LDRD program for support.  This document is released under LA-UR-XX-XXXX.

\bibliography{main}{}
\bibliographystyle{elsarticle-num}
\end{document}